# Direct Observation of Inner and Outer G' Band Double-resonance Raman Scattering in Free Standing Graphene

Zhiqiang Luo[1], Chunxiao Cong[1], Jun Zhang[1], Qihua Xiong[1], Ting Yu[1, 2, 3*]

1. Division of Physics and Applied Physics, School of Physical and Mathematical Sciences, Nanyang Technological University, Singapore, 637371
2. Department of Physics, Faculty of Science, National University of Singapore, Singapore, 117542
3. Graphene Research Centre, National University of Singapore, Singapore, 117542

**ABSTRACT**

In contrary to the widely reported single and symmetric peak feature of G' (or 2D) band in Raman spectrum of graphene, we herein report the observation of splitting in G' band in free standing graphene. Our experimental findings provide a direct and strong support for the previous theoretical prediction that the coexistence of the outer and inner processes in the double-resonance Raman scattering would cause the splitting of G' mode. Our investigation of the influence of trigonal warping effect on the spectral features of G' subbands further verified the theoretical interpretation established on the anisotropic electronic structure of graphene.

**KEYWORDS**

Graphene, Raman Spectroscopy, G'(2D) band, Splitting.

* Corresponding author: Yuting@ntu.edu.sg

Raman scattering has been playing an important role in the study of structural and electronic properties of graphene,[1-8] which is a mono-layer of graphite showing many novel physical properties. Two dominant characteristic Raman features usually present in a Raman spectrum of graphene, the so called G band and G' band.[1] The G band originates from a single resonance process associated with doubly degenerate iTO and LO phonon modes at the Brillouin zone center ($\Gamma$ point), while G' band is associated with two phonon intervalley double resonance (DR) scattering involving iTO phonon near $K$ point.[1] It is widely accepted that the G' band is composed of a single, sharp and symmetric Lorentzian peak for single layer graphene.[2,7,8] However, the theoretical calculation of the G' bands of graphene suggested that the G' band of graphene should be in a split form, since both the outer process in $K\Gamma$ direction and the inner process in $KM$ direction should play significant contribution in DR Raman scattering.[9] Direct observation of the splitting of the D/G' band is thus important for verifying these theoretical results and understanding of the detailed DR Raman scattering process in graphene. In this work, using free standing graphene samples, we observed the splitting of G' band, and further investigated the influence of trigonal warping effect in electronic structure of graphene on these two G'subbands.

The free standing graphene samples were prepared by micromechanical exfoliation of highly ordered pyrolytic graphite (HOPG) onto a $SiO_2$/Si substrate pre-patterned with an array of trenches, which was fabricated by photolithography and reactive ion etching.[10] The Raman spectra were recorded by Renishaw inVia Raman system with excitation lasers of 2.33 eV (532 nm) and 1.58 eV (785 nm), and Jonin-Yvon T64000 Raman system with excitation lasers of 3.49 eV (355 nm) and 1.96 eV (633 nm). The laser power on graphene sample is kept below 1 mW to avoid possible laser-induced heating.

Figure 1 (a) shows an optical image of a typical free standing graphene sample used in this study. For all the free standing graphene samples, there is no noticeable strain in the suspended area as evidenced by the negligible shift of G band compared to those of the non-suspended area[4] (see Raman spectra in Figure 1 (b)). It is interesting to notice that, the G' band of the free standing graphene is obviously asymmetric, while the G' band of the graphene on $SiO_2$/Si substrate appears a broader symmetric peak. The Lorentzian peak fittings of the highly asymmetric G' peak recorded under the excitation of 2.33 eV is displayed in Figure 1 (c). The asymmetric G' band can be well fitted by two Lorentzian components. The splitting of the G' band can be observed in free standing graphene excited with all the laser of energies from near infrared (NIR, 1.58 eV) to ultraviolet (UV, 3.49 eV). The energy dependence of these two sub-bands of the G' band, G'$_1$ at lower frequency and G'$_2$ at higher frequency, were plotted in Figure 1 (d), which have nearly the same slope of 110 cm$^{-1}$/eV.

In most of the previous reports, the observed G' band of graphene on a substrate can always be nicely fitted with one Lorentzian line.[2,7,8] Therefore, it was widely accepted that the G' band was composed of one single, sharp and symmetric peak. Actually, in the intervalley DR Raman scattering of the G' band, there are two possible scattering processes, so called outer process and inner process.[9,11] As illustrated in Figures 2 (a) and (b), the outer process is associated with electron transition in *KΓ* direction, where electrons selectively couple with the iTO phonons along the *KM* direction, while the inner process is associated with electron transition in *KM* direction, where electrons selectively couple with the iTO phonons along the *KΓ* direction.[9,12] Due to the trigonal warping effect[13], the energy dispersion of electron in *KΓ* direction is steeper. The outer process involves the photo-excited electron with smaller momentum, and then phonon with lower frequency, while the inner process involves the photo-excited electron with larger momentum and phonon with higher frequency.[9,12] It would be appropriate to assign the G'$_1$ and G'$_2$ to the outer process and the inner process, respectively. On the basis of peak fitting, the intensity ratio of the inner to the outer processes was estimated of approximately 35% ~ 60%. The smaller weight of the inner process would come from the lower phonon density of states in the *KΓ* direction.[12]

The outer and inner processes of G' band are related to the anisotropic electronic structure of graphene. The trigonal warping effect in graphene is expected to show significant influence on the spectral features associated with the outer and inner process. Actually, there is a sublet change in the full width at half-maximum (FWHM) of G'$_1$ and G'$_2$ at different excitation energies, which should be induced by the trigonal warping effect. As shown in Figure 3 (a), for the excitation of 1.58 eV, the FWHM of the G'$_1$ and G'$_2$ peaks are around 19 cm$^{-1}$. However, for the excitation of 2.33 eV, the FWHM of the G'$_1$ and G'$_2$ peaks are around 15.5 cm$^{-1}$ and 23 cm$^{-1}$, respectively (see Figure 1 (c)). The deviation of the FWHM of the outer and inner process at higher excitation energy should be attributed to the enhanced trigonal warping effect at higher energy level,[13] which causes a flatter edge along the *ΓM* direction and sharper curvature in *KM* direction (see illustration in Figures 2 (c) and (d)). For the photo-electrons at the same energy level, deviance of the change in electron momentum (Δ*k*) during electron scattering is smaller in the flatter edge along the *ΓM* direction, while it becomes larger in the sharper curvature in *KM* direction. Due to momentum conservation during electron-phonon scattering, the change in electron momentum equates the momentum of the phonon ($q = \Delta k$); therefore the smaller deviance of the change in electron momentum at $K\Gamma$ direction will result in smaller FWHM of the Raman peak corresponding to the outer process.

The trigonal warping effect also causes an apparent polarization dependence of the inner to outer processes ratio (see Figure 3). Taking the G' band excited by 1.58 eV laser as an example, when the excitation and detection polarizations are parallel ($\theta = 0°$), the inner to outer processes ratio is about 35%, accompany with the FWHM of the outer and inner process of 19 cm$^{-1}$. Whereas, when the excitation and detection polarizations are orthogonal ($\theta = 90°$), the inner to outer processes ratio is about 45%, accompany with the FWHM of the outer and inner process of 18 cm$^{-1}$ and 20 cm$^{-1}$, respectively. Obviously, as shown in Figures 3 (c) and (d), this polarization dependence becomes more apparent in the UV region. For better understanding of this polarization dependent inner to outer processes ratio, polarized Raman spectra involving of G band and G' band are shown in Figure 4. When $\theta = 0°$, the $I_{G'}/I_G = 10$, $I_{G'1}/I_G = 7.5$, and $I_{G'2}/I_G = 2.5$, respectively. When $\theta = 90°$, the $I_{G'}/I_G = 3.3$, $I_{G'1}/I_G = 2.1$, and $I_{G'2}/I_G = 1.2$, respectively. It is well known that the G band intensity of graphene is isotropic; however, due to the inhomogeneous optical absorption and emission in DR Raman scattering process associated with the G' band, the G' band intensity of graphene is anisotropic: $I_{G'}(\theta) = [2\cos^2(\theta)+1]*I_0/3$, where $\theta$ is the angle between the polarization of the analyzer and the polarization of the incident laser, and the $I_0$ is a maximum Raman intensity for the G' band when $\theta = 0°$.[14] The $(I_{G'}/I_G)_{\theta = 90°}/(I_{G'}/I_G)_{\theta = 0°} = 3.3/10 = 1/3$ from our measurement is in good agreement with the calculation results.[14] The $I_{G'}(\theta)$ is the intensity integrated over the whole Brillouin Zone, involving the contribution from both the outer and the inner processes.[14] The polarization dependence of the intensity of the separated G' subbands corresponding to the outer and inner processes should not simply follow the $I_{G'}(\theta)$.

The absorption probability in the light absorption process as a function of angle around the K point shows significant difference for the outer and inner processes,[15] which should result in different polarization dependence for G' subbands. The electron states around $KM$ direction with sharper curvature have higher absorption probability accompany with wider angle distribution.[14,15] In the inner process, wider angle distribution of optical absorption at $KM$ direction should cause smaller weight in the projection of the scattered light (P$s$) along the polarization direction of incident light (P$_L$), and therefore the $(I_{G'2}/I_G)_{\theta = 90°}/(I_{G'2}/I_G)_{\theta =0°}$ should be larger than 1/3. On the contrary, the angle distribution of optical absorption at $K\Gamma$ direction with flatter edge is smaller in the outer process, resulting in larger weight in the projection of P*s* along P*L*, and therefore the $(I_{G'1}/I_G)_{\theta = 90°}/(I_{G'1}/I_G)_{\theta = 0°}$ should be smaller than 1/3. In Figure 4, the $(I_{G'1}/I_G)_{\theta = 90°}/(I_{G'1}/I_G)_{\theta = 0°} = 0.28$, and $(I_{G'2}/I_G)_{\theta = 90°}/(I_{G'2}/I_G)_{\theta = 0°} = 0.48$, which well fits the above theoretical picture. Therefore, it is clear that the polarization dependence of the inner to outer processes ratio results from the trigonal warping effect, which well supports the interpretation of our observed G' splitting on basis of the anisotropic electronic structure of graphene.[9]

In Summary, the two well fitted Lorentzian lines in our observed asymmetric G' band of the free standing graphene should be a direct evidence of the coexistence of both outer process and inner process in the DR Raman scattering. The trigonal warping effect in electronic structure of graphene shows significant influence on the FWHM and polarization dependence of the subbands associated with the outer and inner processes, which well supports the theoretical interpretation established on the anisotropic electronic structure of graphene. This interesting observation experimentally reveals the detailed Raman scattering process corresponding to the G' band in graphene.

**Acknowledgements**

Yu acknowledges the support by Singapore National Research Foundation under NRF RF Award No. NRFRF2010-07 and MOE Tier 2 MOE2009-T2-1-037.

**Figure captions**

**Figure 1** (Color online) (a) An optical image of the free standing graphene sample; (b) Raman spectra of the free standing graphene and the graphene on $SiO_2$/Si substrate (excitation energy is 2.33 eV); (c) Lorentzian fitting of G' band of the free standing graphene; (d) Energy dispersion of the G' subbands of the free standing graphene.

**Figure 2** (Color online) The outer (a) and inner (b) scattering processes of G' band. The influence of trigonal warping effect on the outer and inner processes is illustrated in (c) for low energy level and (d) for high energy level. The outer scattering process is in black lines, and the inner scattering process is in red lines.

**Figure 3** (Color online) Polarized G' band Raman spectra of the free standing graphene. The incident laser beam is in a fixed polarization and the polarization angles of the analyzer respect to incident polarization are $0^o$ and $90^o$. The excitation energies are 1.58 eV for (a) and (b), and 3.49 eV for (c) and (d), respectively.

**Figure 4** (Color online) Polarized Raman spectra of the free standing graphene under excitation of 1.96 eV.

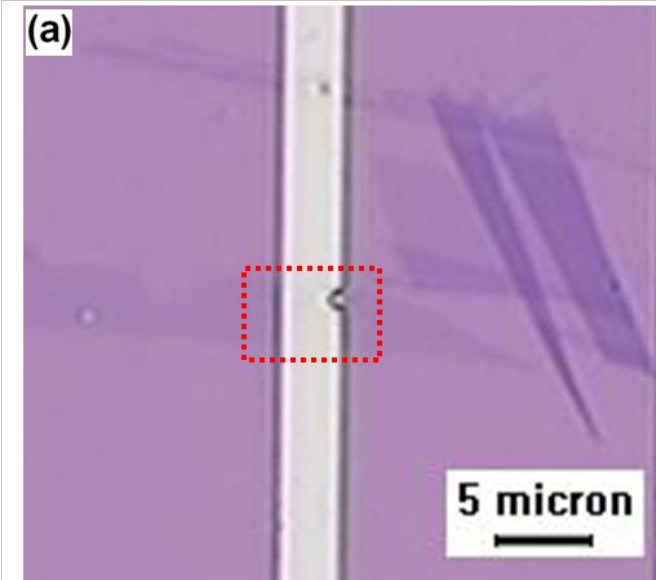 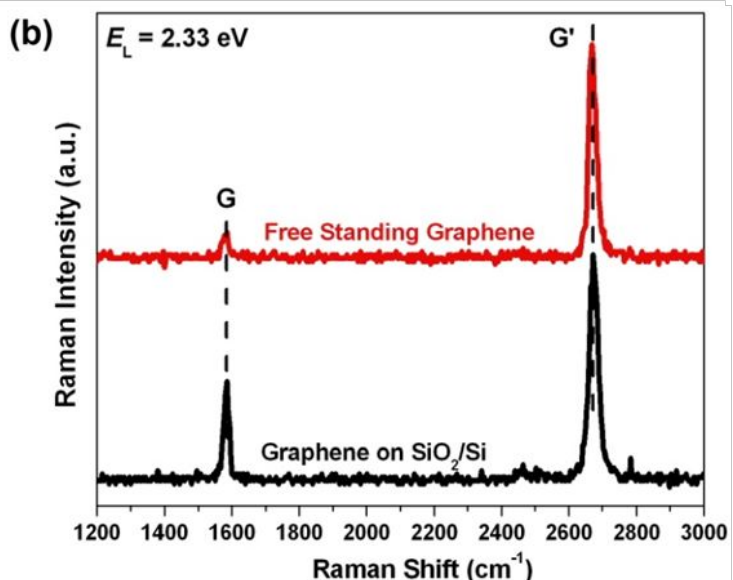
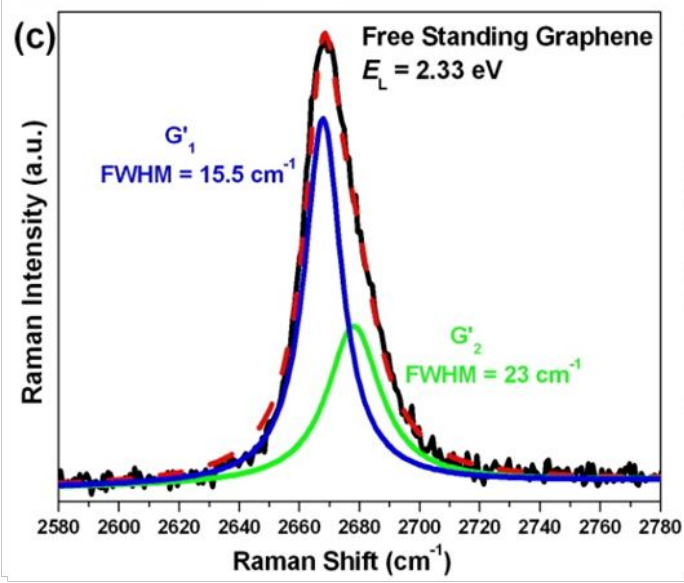 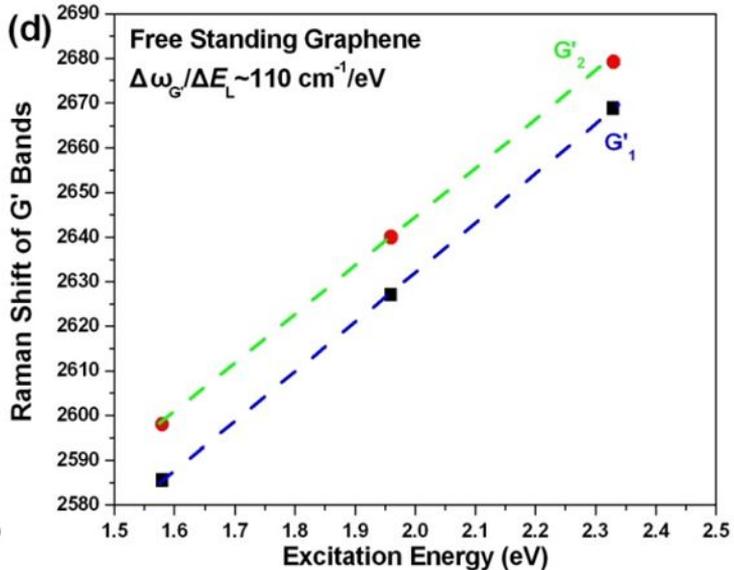

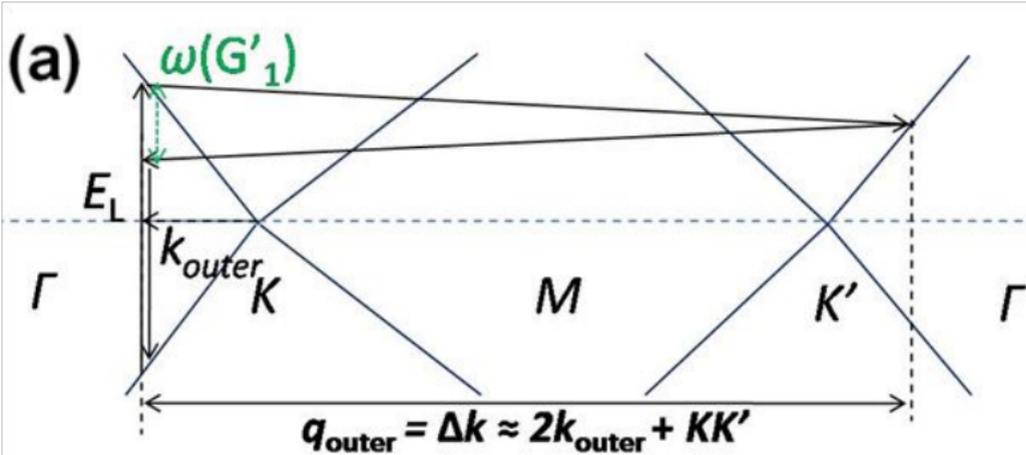
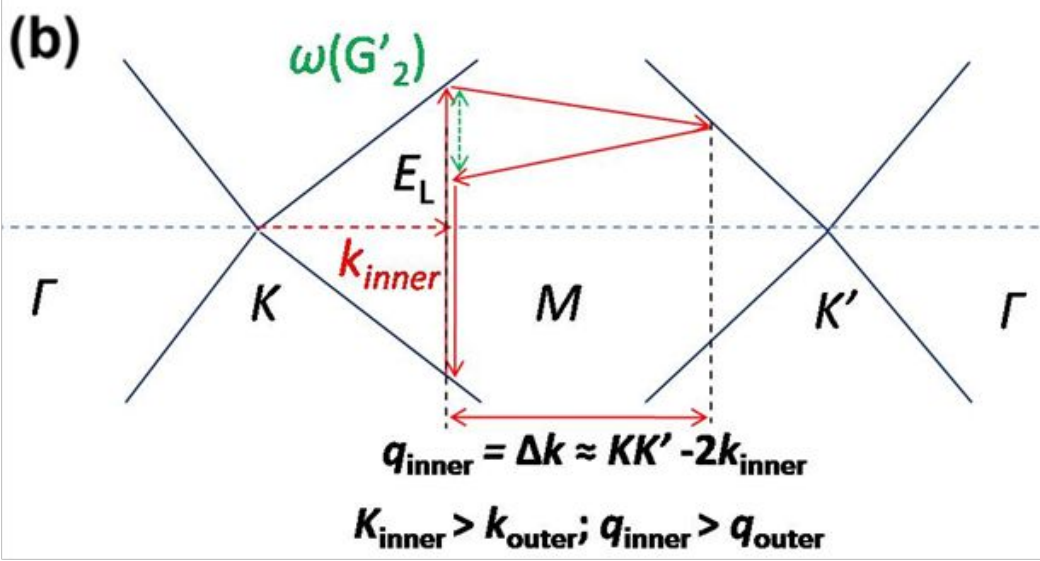
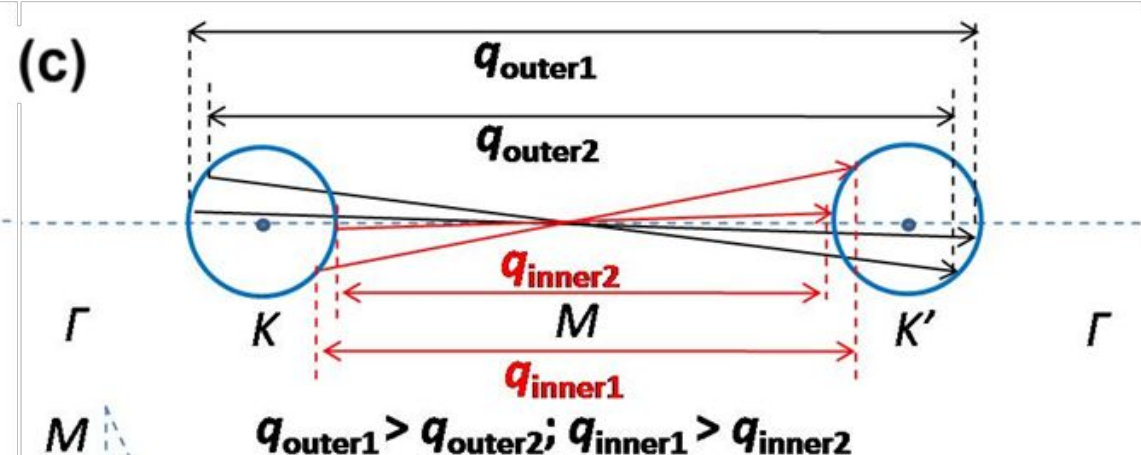
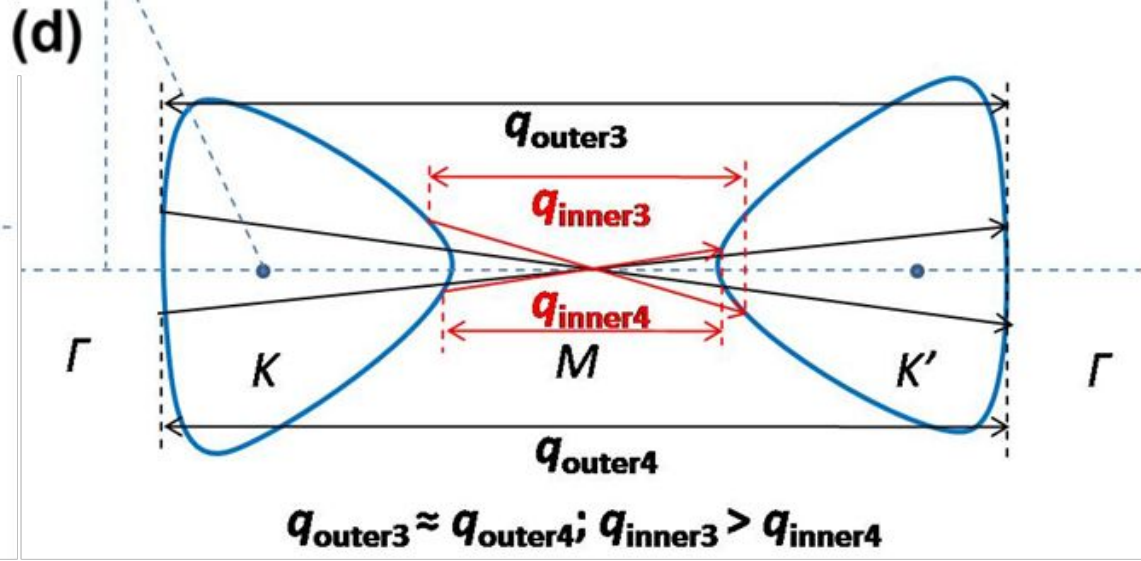

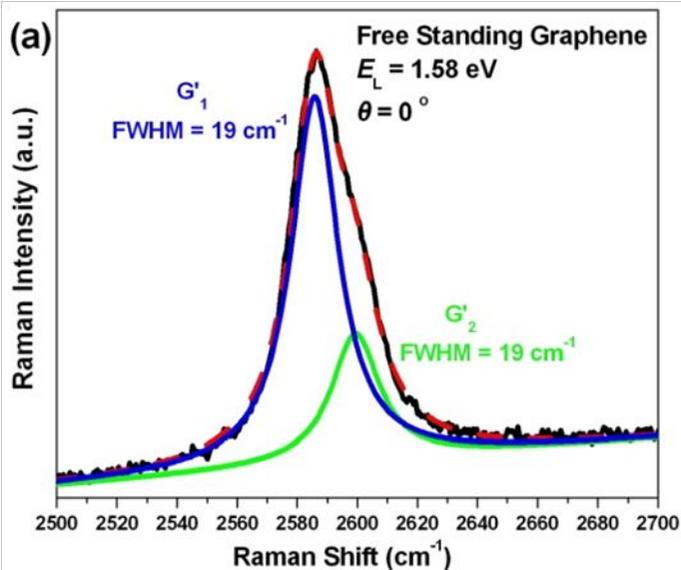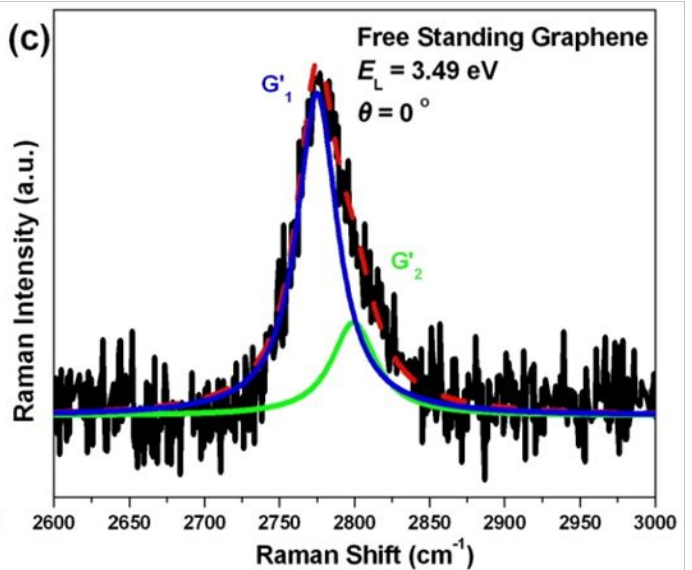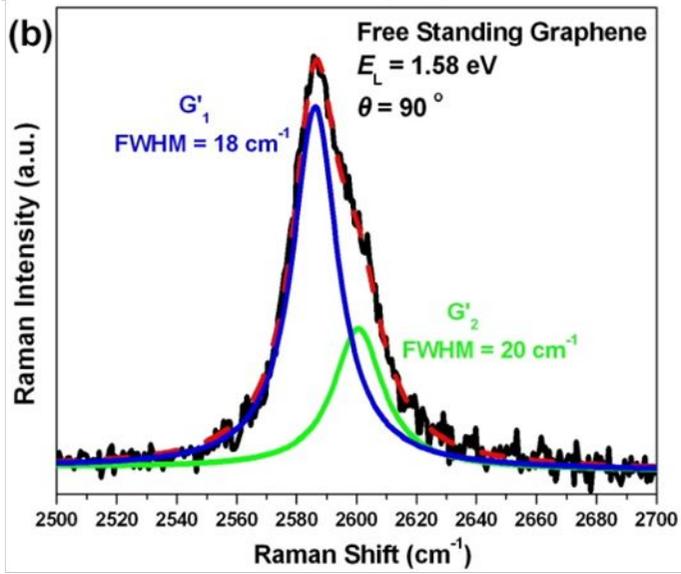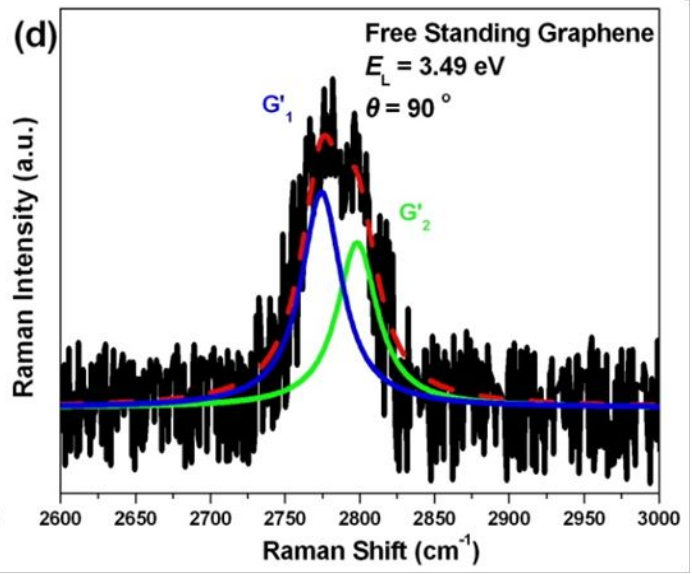

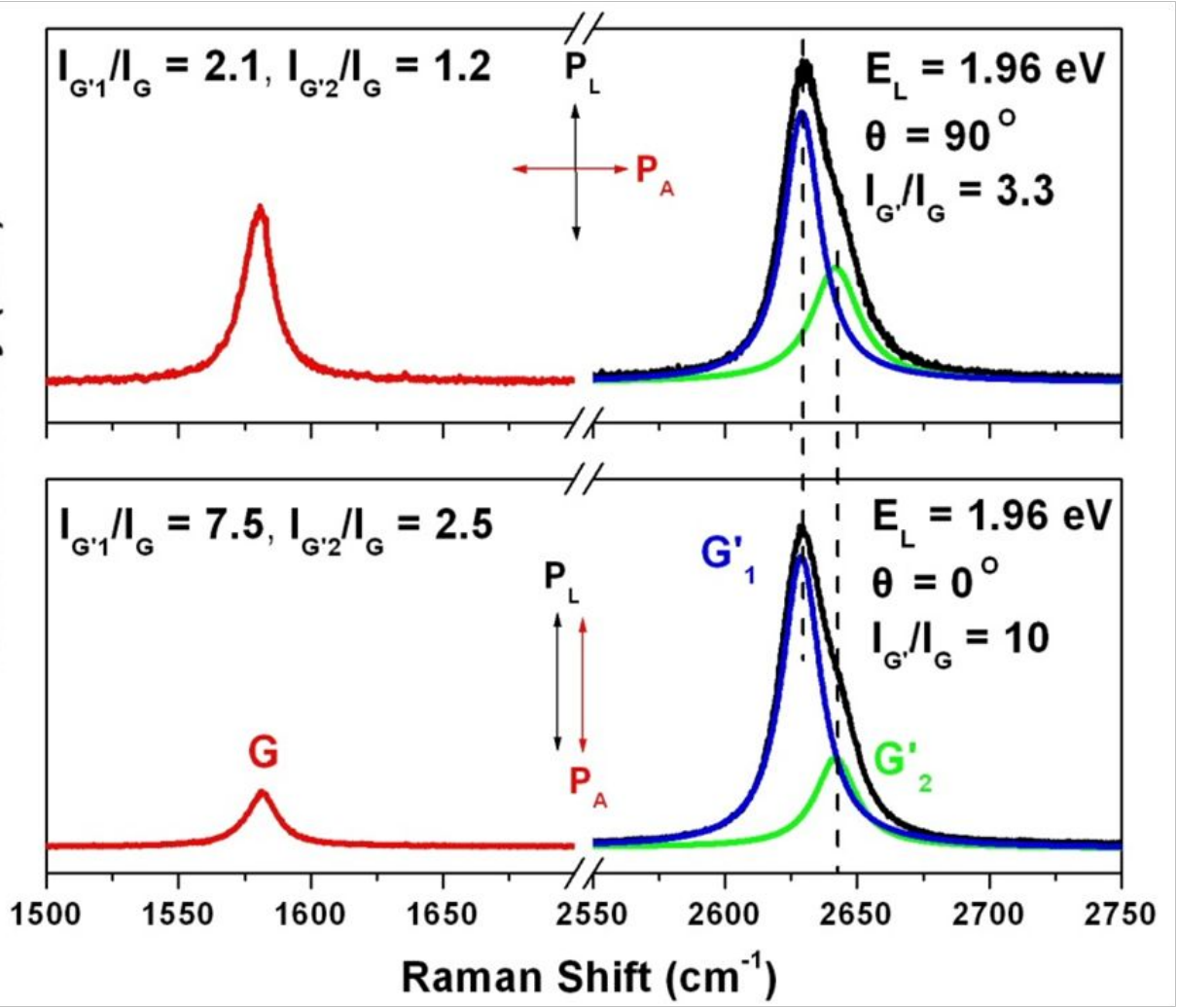